\documentclass{kluwerhyf}    % Specifies the document style.

\newdisplay{guess}{Conjecture}

\begin{document}                                                                                   
\begin{article}
\begin{opening}         
\title{Beaming Effects in GRBs and Orphan Afterglows} 
\author{Y. F. \surname{Huang}}  
\institute{Department of Astronomy, Nanjing University, Nanjing 210093, China}
\author{T. \surname{Lu}}  
\institute{Purple Mountain Observatory, CAS, Nanjing 210008, China}
\author{K. S. \surname{Cheng}}  
\institute{Department of Physics, The University of Hong Kong, Hong Kong, China}
\runningauthor{Y. F. Huang, T. Lu \& K. S. Cheng}
\runningtitle{Beaming in GRB afterglows}
\date{June 1, 2004}

\begin{abstract}
The overall dynamical evolution and radiation mechanism of $\gamma$-ray burst 
jets are briefly introduced. Various interesting topics concerning beaming in 
$\gamma$-ray bursts are discussed, including jet structures, orphan afterglows 
and cylindrical jets. The possible connection between $\gamma$-ray bursts and 
neutron stars is also addressed. 
\end{abstract}

\end{opening}           

\section{Introduction}  
                    % Produces section heading.  Lower-level
                    % sections are begun with similar 
                    % \subsection and \subsubsection commands.

The discovery of $\gamma$-ray burst (GRB) afterglows in 1997, triggered 
by the famous Italian-Dutch BeppoSAX satellite, definitely shows that
most, if not all, long GRBs are of cosmological origin. The so called 
``fireball model'' is strongly favored theoretically. In this standard model, 
the GRB fireball is  assumed to be isotropic. However, as early as in 
1997, Rhoads (1997) has already suggested that GRB outflows may be highly 
collimated. In the beaming case, as the ultra-relativistic jet decelerates, 
it will expand laterally at approximately co-moving sound speed. Naturally, 
photons are emitted into larger and larger solid angle. As the result, 
an obvious break should be observed in the multi-band afterglow light curves.
The break time is determined by $\gamma \sim 1/\theta$, where $\gamma$ is 
the bulk Lorentz factor of the jet and $\theta$ is its half opening angle.

Observationally, the jet hypothesis gains some support soon in 1997.  
The $\gamma$-ray energy release of GRB 971214, if isotropic, is as 
large as $\sim 0.17 M_\odot c^2$, well beyond the energy scope of 
a stellar object. Similar difficulty also exists in many other 
examples, such as GRBs 980703 ($\sim 0.06 M_\odot c^2$), 990123 ($\sim 
1.9 M_\odot c^2$), 990510 ($\sim 0.16 M_\odot c^2$), 991208 ($\sim
0.07 M_\odot c^2$), 991216 ($\sim 0.38 M_\odot c^2$), 000131 ($\sim 
0.6 M_\odot c^2$), 000926 ($\sim 0.15 M_\odot c^2$), 010222 ($\sim 
0.3 M_\odot c^2$), and 020813 ($\sim 0.55 M_\odot c^2$). In all these
cases, emission should be highly collimated, so that the true energy
release can be reduced to $\sim 10^{50}$ --- $10^{51}$ ergs, within 
the energy output of a stellar object. 

Also it is very interesting that light curve breaks do have been
observed in a few afterglows, for example, in GRBs 990123,
990510, 991216, 000301C, 000926, 010222, 011121, 020124, 020813,
030226, and 030329. Such breaks have been widely regarded as being due
to jet effect. In a few other cases (GRBs 980326, 980519, 990705,
991208, 000911, 001007, 020405), although no breaks were observed,
the light curves are still abnormal since the afterglows decay
quite steeply ($\sim t^{-2}$). Such rapid fading of optical
afterglows has also been argued as evidence for collimation
(Huang, Dai \& Lu 2000b).

Beaming is an interesting topic in the field of GRBs. There are
many researches concerning it, and many interesting results have been
revealed. For example, Frail et al. (2001) suggested that the 
intrinsic energy releases of GRBs, after correction for the beaming
angle, are strikingly clustered around $5 \times 10^{50}$ ergs. 
Recently, it is also discovered that a GRB jet should be highly
structured, but not homogeneous. 

In this article, we mainly discuss beaming effects in GRB afterglows. 
The dynamics and radiation mechanism will be described in Section 2.
Structures of jets are then introduced in Section 3. The possible 
existence of cylindrical jets is addressed in Section 4. Section 5
is about orphan afterglows, and Section 6 investigates the 
possibility that the launch of a GRB jet may be associated with
the kick of a high speed neutron star. The final section is a 
brief discussion. 

\section{Dynamics and Radiation}

After producing the main burst via internal shocks at a radius 
about $10^{13}$ cm, the GRB ejecta continues to expand ultra-relativistically
in the circum-burst medium. The external shock occurs when the swept-up 
medium mass, $m$, exceeds $M_{\rm ej}/\eta$, where $M_{\rm ej}$ is the 
the initial mass of the ejecta and $\eta$ is the initial value of the 
Lorentz factor $\gamma$. Afterglows are produced by the shock-accelerated 
electrons. Denoting the radius of the external shock as $R$, the observer's
time as $t$, the medium number density as $n$, then the overall evolution of a 
GRB jet can be conveniently described as (Huang et al. 1999, 2000a, b, c),
\begin{equation}
\label{drdt1}
\frac{d R}{d t} = \beta c \gamma (\gamma + \sqrt{\gamma^2 - 1}),
\end{equation}
\begin{equation}
\label{dmdr2}
\frac{d m}{d R} = 2 \pi R^2 (1 - \cos \theta) n m_{\rm p},
\end{equation}
\begin{equation}
\label{dthdt3}
\frac{d \theta}{d t} = \frac{c_{\rm s} (\gamma + \sqrt{\gamma^2 - 1})}{R},
\end{equation}
\begin{equation}
\label{dgdm4}
\frac{d \gamma}{d m} = - \frac{\gamma^2 - 1}
       {M_{\rm ej} + \epsilon m + 2 ( 1 - \epsilon) \gamma m}, 
\end{equation}
where $\beta = \sqrt{\gamma^2-1}/\gamma$, and $\epsilon$ is the radiative efficiency.
$c_{\rm s}$ is the co-moving sound speed, which can be further expressed as,
\begin{equation}
\label{cs5}
c_{\rm s}^2 = \hat{\gamma} (\hat{\gamma} - 1) (\gamma - 1) 
	      \frac{1}{1 + \hat{\gamma}(\gamma - 1)} c^2 , 
\end{equation}
where $\hat{\gamma} \approx (4 \gamma + 1)/(3 \gamma)$ is the adiabatic 
index. 

This dynamical model has the advantage that it applies in both the 
ultra-relativistic and the non-relativistic phases, and that it describes the lateral
expansion in an accurate way.

Synchrotron radiation is the main emission mechanism. To make our calculation 
appropriate even in the deep Newtonian phase (Huang \& Cheng 2003), we assume 
that the shock-accelerated electrons distribute according to their kinetic 
energy as (Huang \& Cheng 2003), 
\begin{equation}
\label{eq8}
\frac{d N_{\rm e}'}{d \gamma_{\rm e}} \propto (\gamma_{\rm e} - 1)^{-p} , 
\;\;\; (\gamma_{\rm e,min} \leq \gamma_{\rm e} \leq \gamma_{\rm e,max}),
\end{equation}
where $\gamma_{\rm e}$ is the thermal Lorentz factor of electrons. Assuming that there
is an equi-partition between the proton energy density, the magnetic energy
density, and the electron energy density as well, it will then be relatively
easy to calculate the afterglows by considering synchrotron radiation. Note 
that the equal-time-surface effect should be taken into account in calculations. 
Examples of such calculations have been given in Huang \& Cheng (2003). 

\vspace{1cm}

\section{Jet Structure}

The simplest jet model involves a homogeneous conical outflow. 
Recently it was realized by more and more authors that GRB jets
may have complicate structures. Basically there are three kinds
of structured jets: two-component jets (Berger et al. 2003), 
Gaussian jets (where the
energy per unit solid angle depends as a Gaussian function on the
angular distance from the axis), and power-law jets (where the
energy density profile is a power-law function) (M\'esz\'aros,
Rees \& Wijers 1998; Dai \& Gou 2001; Zhang \& M\'esz\'aros 2002).
Generally, the structured jet models have the potential of explaining normal
GRBs, X-ray rich GRBs, and X-ray flashes in a uniform picture
(Huang et al. 2004; Zhang et al. 2004).

Although the profile functions of Gaussian jets and power-law jets seem
quite simple, their afterglows are in fact not easy to calculate, especially
when the lateral expansion and the equal-time-surface effect are considered. 
The two-component jet model is relatively simple in these aspects. 
A two-component jet has two components: a narrow but ultra-relativistic
outflow (with Lorentz factor typical of normal GRB fireballs, i.e. $\gamma 
\geq 100$ --- 1000), and a wide but mildly 
relativistic ejecta (with $1 \ll \gamma \ll 100$). These two components are 
usually assumed to be coaxial. At first glance, the two-component jet model
seems to be quite coarse, but interestingly enough, 
it gains some support from numerical simulations of the collapse of massive stars 
(Zhang et al. 2003). Additionally, Berger et al. (2003) found that the model 
can give a perfect explanation to the multiband observations of the 
famous GRB 030329. In their explanation, the gamma-ray and early afterglow 
emission of GRB 030329 come from the narrow component, while the radio and optical
afterglows beyond 1.5 days are produced by the wide component. 

\begin{figure} % figuur 1
\vspace{12pc}
\includegraphics{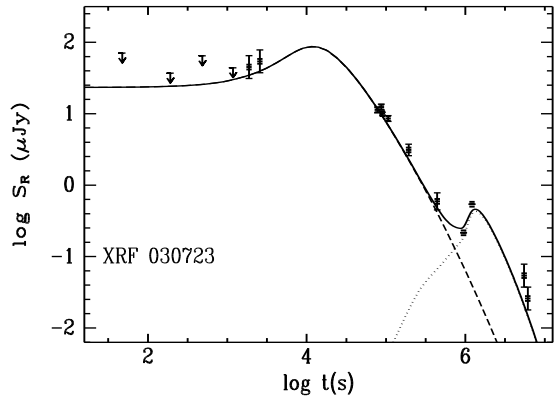}
\caption[]{Fit to the optical afterglow of XRF 030723 by Huang et al. (2004), 
using the two-component jet model. The dashed line corresponds to emission from the 
wide component, and the dotted line is for the narrow component. The solid line 
illustrates the total light curve. The observer is assumed to be off-axis.}
\label{FigXRF}
\end{figure}

In a recent study, Huang et al. (2004) further proposed that the optical afterglow
light curve of X-ray Flash (XRF) 030723 can also be well fit by the 
simple two-component model. To re-produce the rebrightening of the afterglow 
of XRF 030723, Huang et al. (2004) assumed that the observer is off-axis, and
that the intrinsic energy of the wide component is less than that of the narrow 
component. Figure.~\ref{FigXRF} illustrates the result of their fitting. 
Anyway, it should be noted that the rebrightening in this event can also
be explained by an underlying supernova (Fynbo et al. 2004; Tominaga et al., 2004).

\section{Cylindrical Jets}

\begin{figure} % figuur 1
\vspace{11pc}
\includegraphics{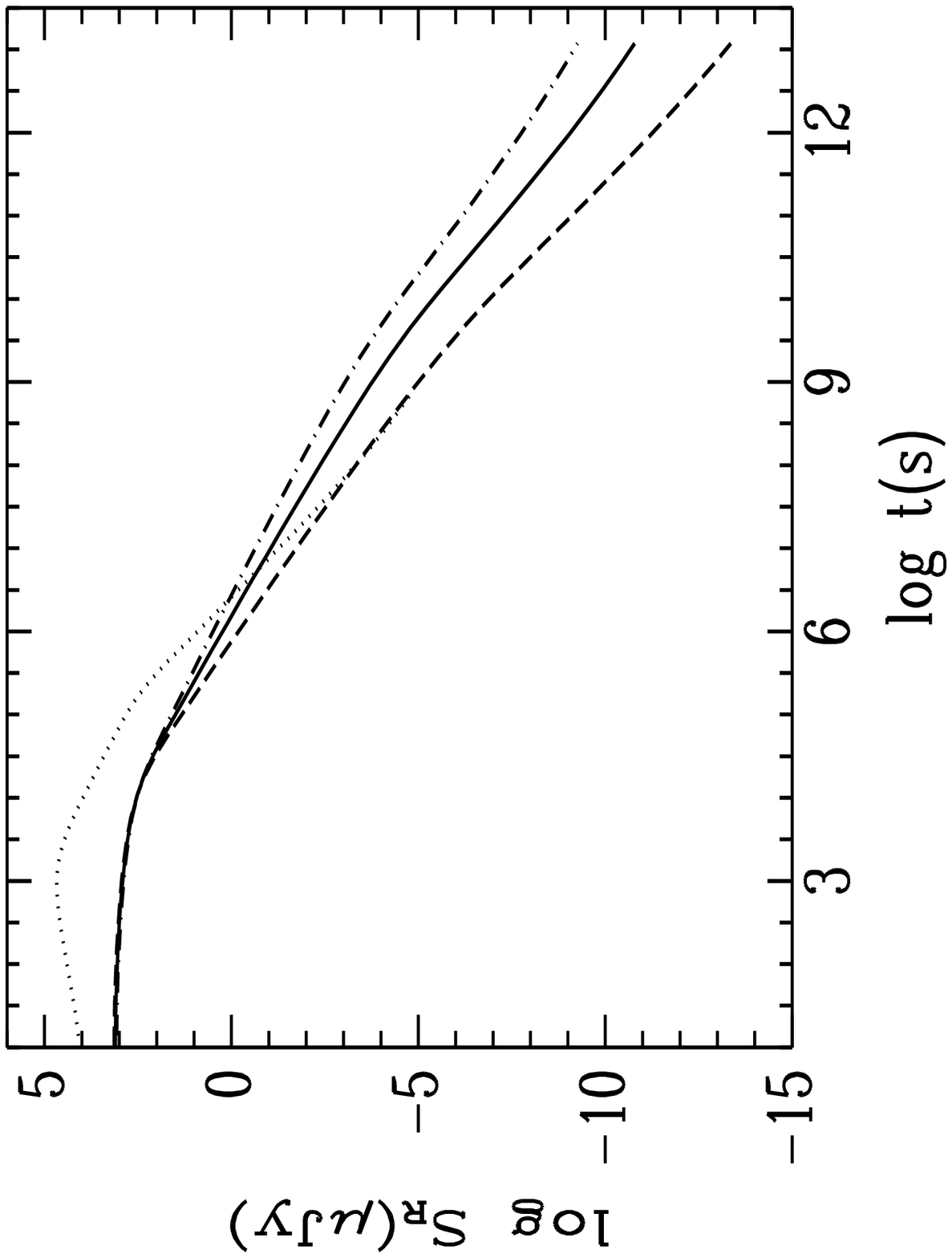}
\includegraphics{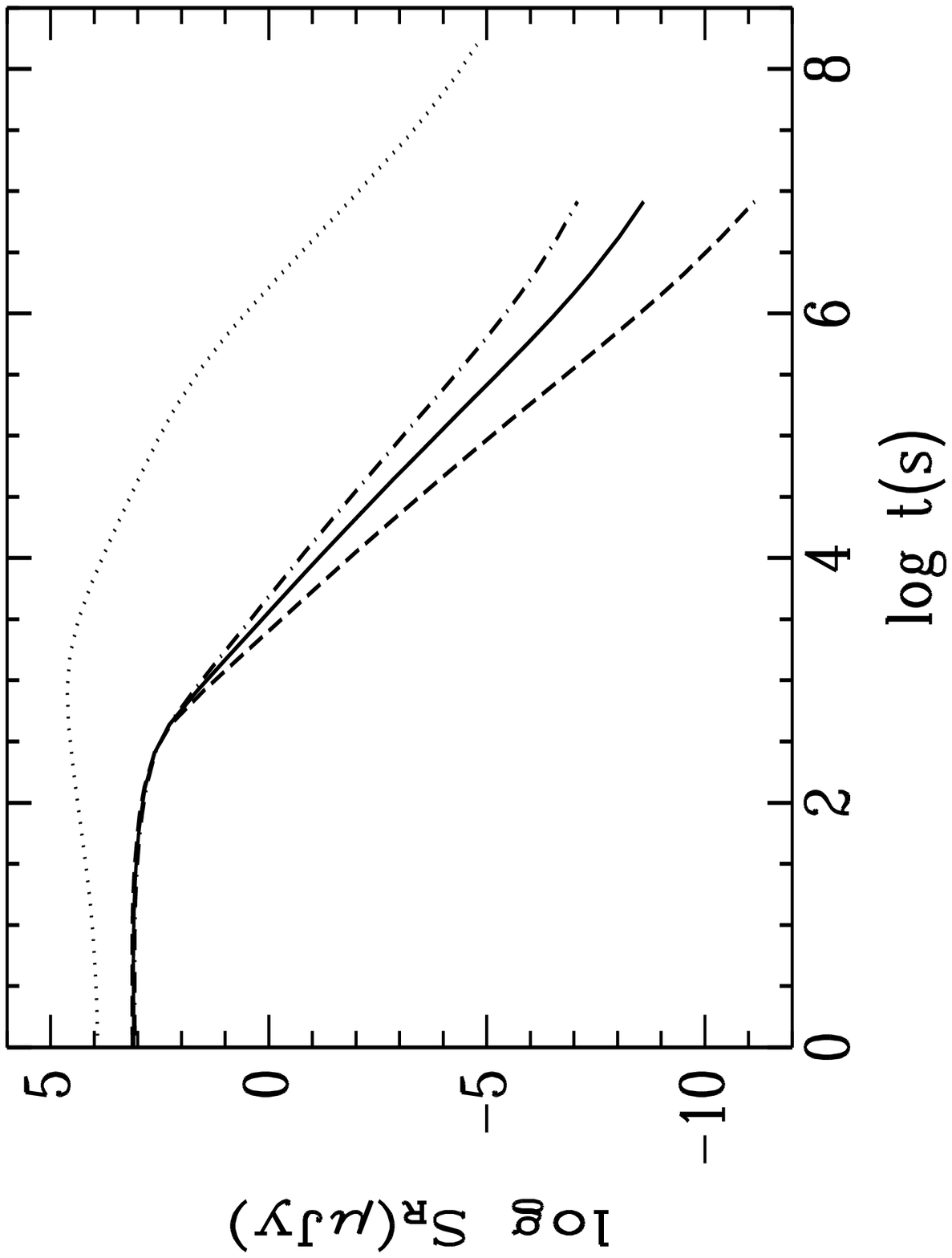}
\caption[]{R-band afterglows from cylindrical jets without (left panel) and with
(right panel) lateral expansion (Cheng, Huang \& Lu 2001). The dashed, solid and 
dash-dotted lines in each panel correspond to $p=3, 2.5$ and 2.2 respectively. 
The dotted lines correspond to conical jets with $p=2.5$. }
\label{CynLight}
\end{figure}

Usually GRB jets are assumed to be conical outflows. However, Cheng, Huang 
and Lu (2001)  have suggested that the relativistic outflows in GRBs might 
also be cylindrical. They have studied afterglows of cylindrical jet detailedly. 
If a cylindrical jet does not expand laterally, it will remain in the 
ultra-relativistic phase for a very long period (typically longer than $10^9$ s). 
The afterglow usually decays like $S_{\nu} \propto t^{-p/2}$, where $p$ is 
the power-law index of the electron distribution. On the other hand,
if the cylindrical jet expands laterally, it will enter the Newtonian 
phase quickly. In this case, the afterglow light curve evolves from 
$S_{\nu} \propto t^{-p}$ to $S_{\nu} \propto t^{-(15p-21)/10}$. 
As the example, Figure~\ref{CynLight} illustrates the optical afterglow light 
curves of some cylindrical jets. 

\begin{figure} % figuur 1
\vspace{11pc}
\includegraphics{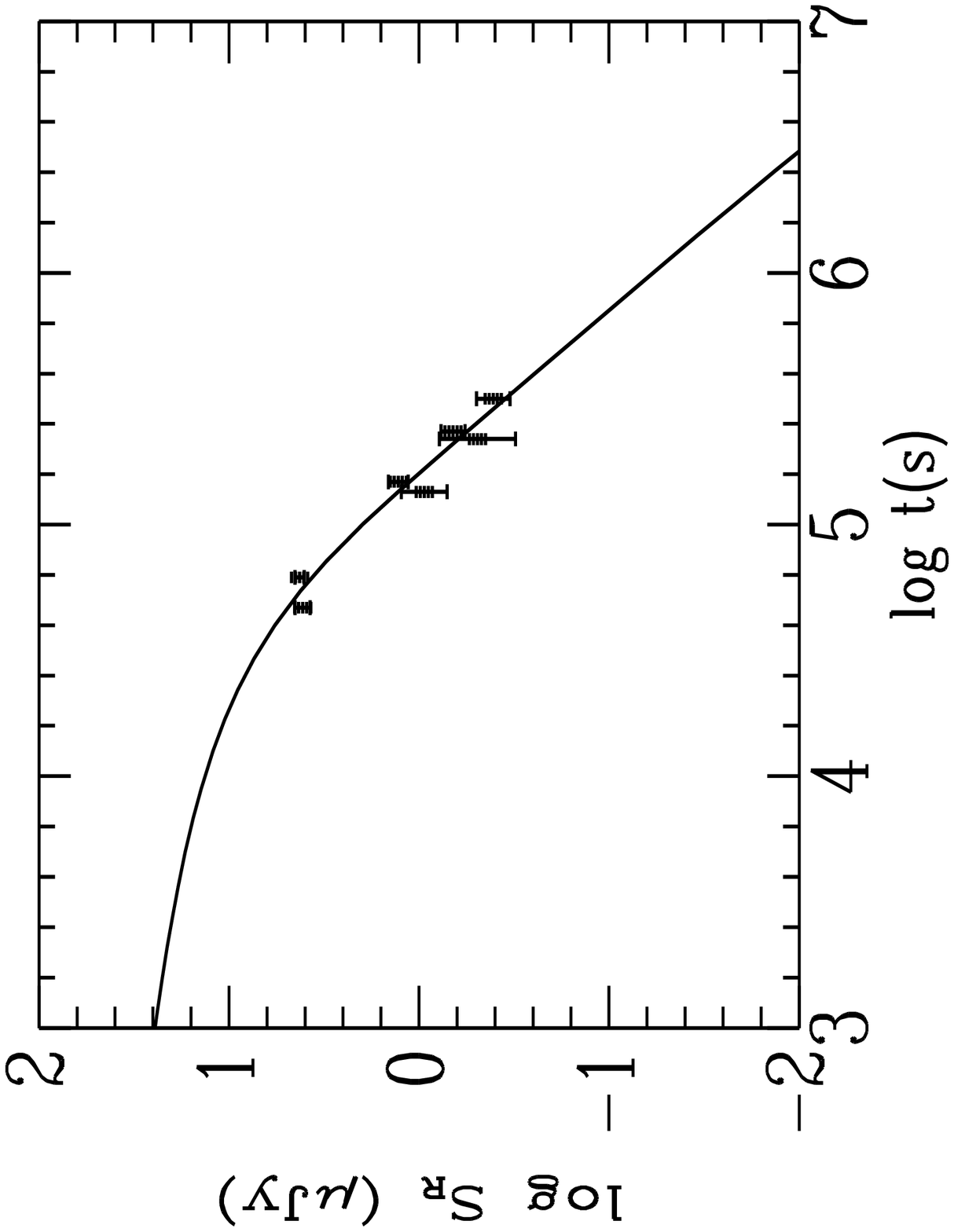}
\includegraphics{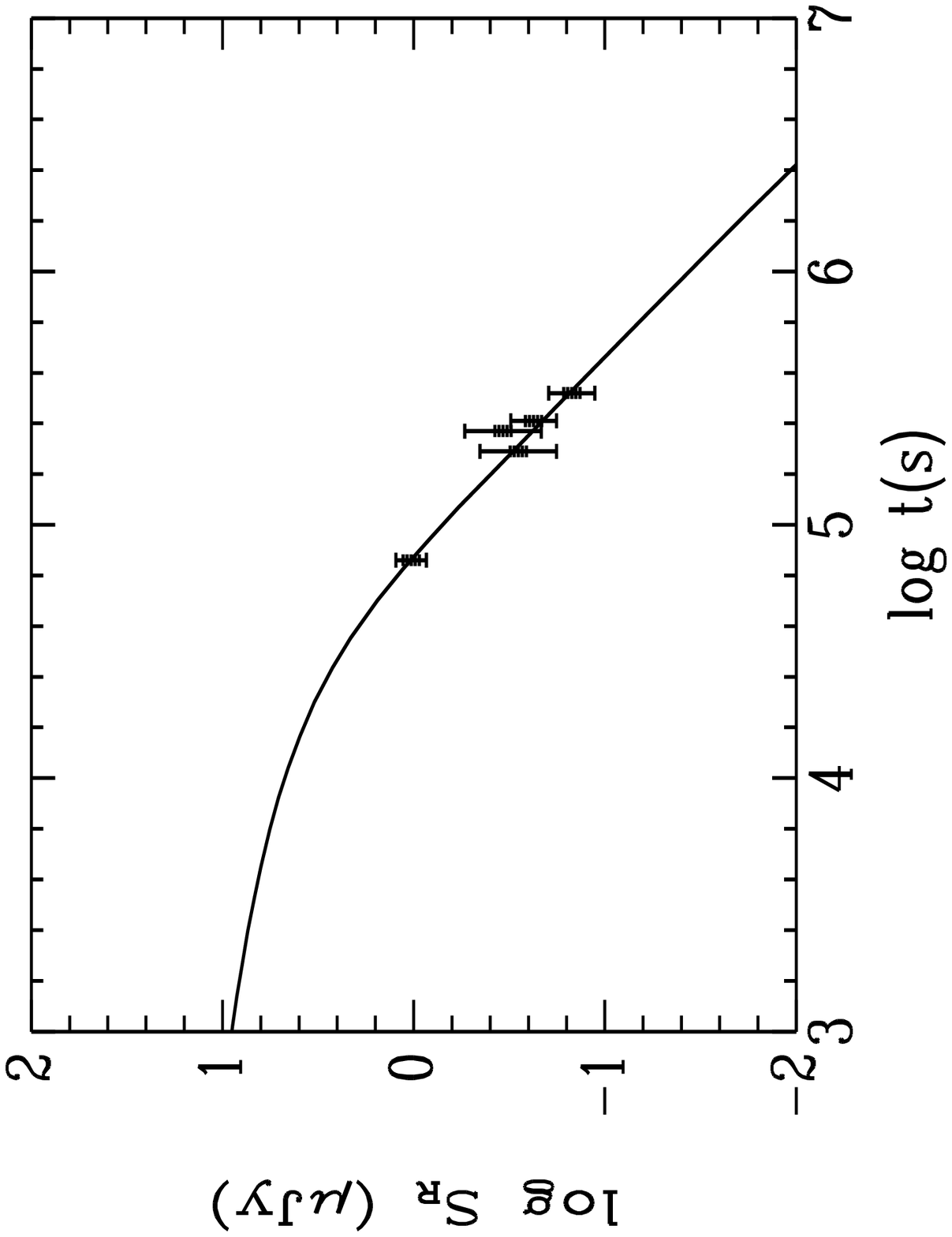}
\caption[]{Fitting to the R-band afterglow light curves of GRB 971214 (left panel) and
GRB 980329 (right panel) by using the cylindrical jet model (Huang et al. 2002b).  }
\label{CynFit}
\end{figure}

Huang et al. (2002b) specially pointed out that for a cylindrical jet without 
lateral expansion, the afterglow light curve will become $S_{\nu} \propto t^{-1}$
--- $t^{-1.3}$ if taking $p=2.0$ --- 2.6 . Observationally, the decay of optical 
afterglows from many GRBs, such as GRBs 970508, 971214, 980329 and 980703, is 
in this range. In the most popular explanation, these GRBs are thought to 
be produced by isotropic fireballs. However, we should not omit the possibility 
that these events may in fact be due to cylindrical jets, as suggested 
by Huang et al. (2002b). Figure~\ref{CynFit} shows that the cylindrical jet model
can fit the afterglows of these events perfectly. 

The concept of cylindrical jets has gained support observationally in fields 
other than GRBs. For example, it has long been found that jets in many radio 
galaxies are cylindrical, i.e. they maintain constant cross sections on large scales. 
Jets in many Herbig-Haro (HH) objects are also cylindrical (e.g., Ray et al. 1996). 
In fact, observations have indicated clearly that HH jets are initially poorly 
focused, but are then asymptotically collimated into cylinders (Ray et al. 1996).

\begin{figure} % figuur 1
\vspace{12pc}
\includegraphics{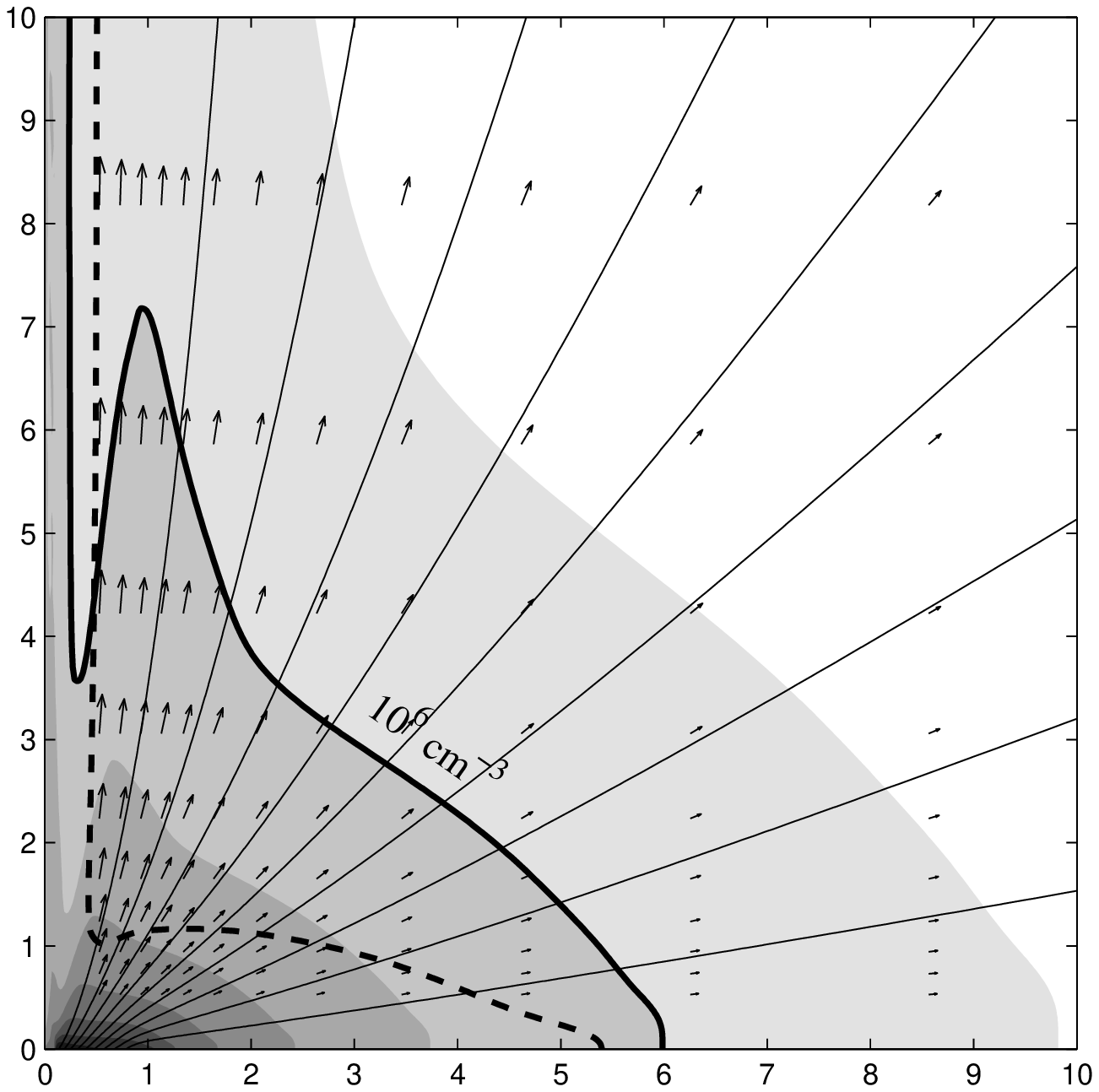}
\includegraphics{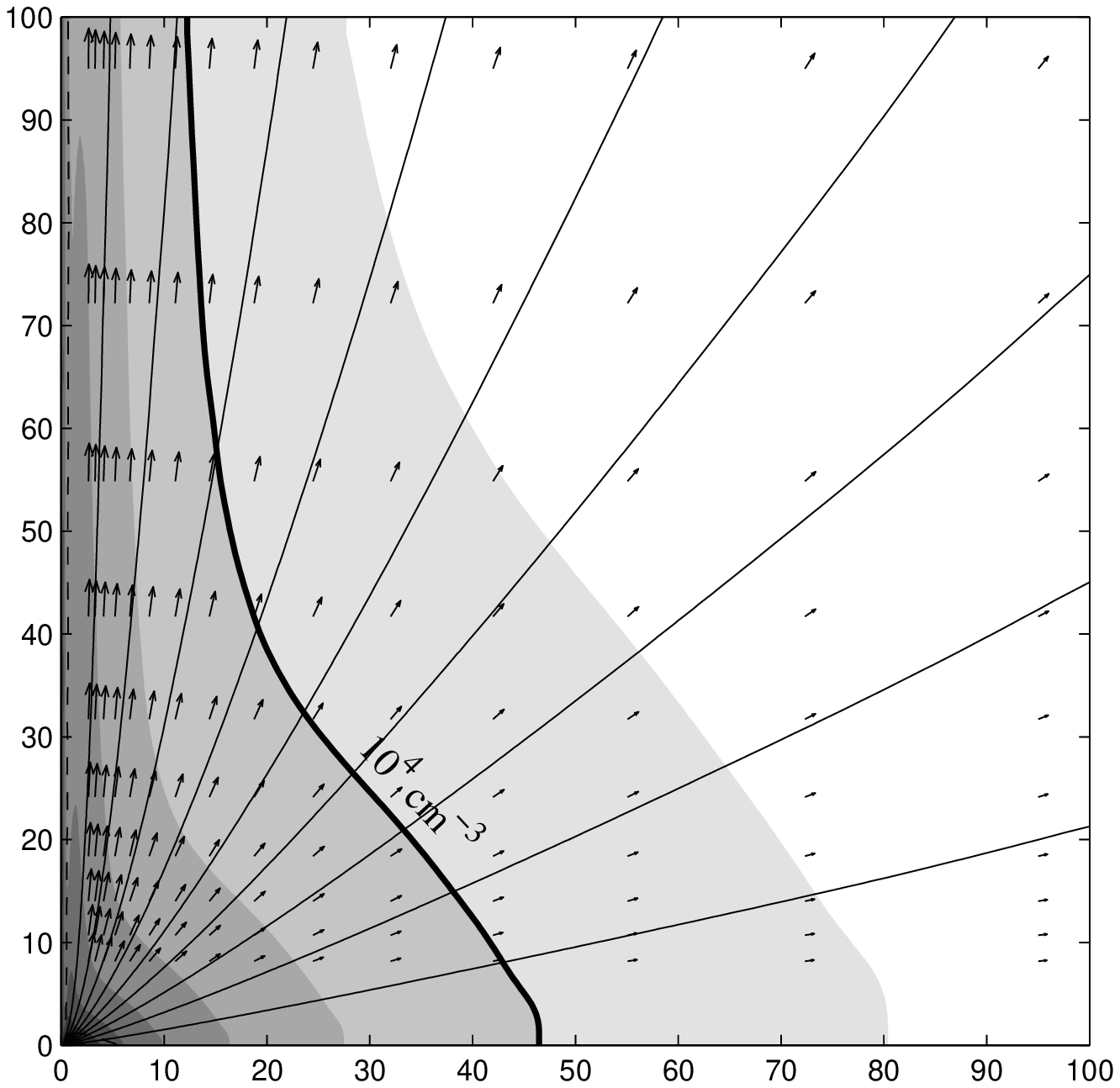}
\caption[]{Numerical results from the calculation of MHD jets launched by a
stellar accretion disk (Krasnopolsky et al. 2003). Shown are streamlines (light solid 
lines) and isodensity contours (heavy solid lines and shades). The arrows are for 
poloidal velocity vectors, with length proportional to the speed. In the Left panel
the jet is plot on 10 AU scale, and in the Right panel the jet is plot on 100 AU 
scale. It is clearly seen from the isodensity contours that the jet has a cylindrical
shape. }
\label{Cylinder}
\end{figure}

Theoretically, it is striking that cylindrical jets can be naturally produced in 
black hole-accrection disk systems (Shu et al. 1995; Krasnopolsky et al. 2003; 
Vlahakis \& K\"onigl 2003a, b; Fendt \& Ouyed 2004). The collimation is mainly 
due to magnetic forces. It is well known that the poloidal component of a dipolar 
magnetic field decays as $B_{\rm P} \propto r^{-3}$, where $r$ is the distance from 
the central object. It is also known that the motion of matter along poloidal 
magnetic field lines will unavoidably induce a strong toroidal field component, 
which decays as $B_{\rm T} \propto r^{-1}$ (Fendt \& Ouyed 2004). So, a 
magnetohydrodynamic (MHD) jet is asymptotically dominated by the toroidal magnetic
field. This toroidal field exerts an inward force on the MHD jet through ``hoop 
stress'', which provides the collimation. Numerous numerical results have shown 
that MHD jets are conical initiallly during the acceleration phase, but their 
half opening angles are turning smaller and smaller, until finally the outflows 
become cylindrical. Figure~\ref{Cylinder} shows examplar numerical results by Krasnopolsky
et al. (2003). Of course, in the cases of GRBs, which are thought to occur in 
star forming regions, strong gradients in density might also play a role in 
collimating the jets.

\section{Orphan Afterglows}

If GRBs are really due to beamed ejecta, then the high-energy burst can be observed 
only when the observer is on-axis. However, in the off-axis case, since the afterglow 
emission is less beamed, it is still possible that the ejecta may be detected in 
optical and radio bands. These afterglows are called orphan afterglows, since
they are not associated with any known GRBs. Rhoads (1997) has pointed out that
the ratio of orphan afterglows with respect to GRBs can potentially give a measure 
of the beaming angle of GRB jets. 

However, Huang et al. (2002a) argued that the detection of orphan afterglows does
not necessarily mean that GRBs are jetted. They argued that in the isotropic fireball
model, there should exist many failed GRBs, i.e., fireballs with initial Lorentz 
factor $1 \ll \eta \ll 100$ --- 1000. These fireballs cannot produce GRBs successfully. 
Huang et al. called them failed GRBs (FGRBs), although they sometimes are also called 
dirty fireballs (Dermer et al. 1999).  It is obvious that FGRBs can also produce 
orphan afterglows. Huang et al. (2002a) thus suggest that when an orphan afterglow
is observed, it should be monitored carefully for a relative long period so that its
origin can be clarified. It can be used to estimate the beaming angle of GRBs only 
when we know for sure that it really comes from a jetted but off-axis GRB. 

\section{GRB Jets and Neutron Star Kicks}

Since the discovery of afterglows in 1997, great progresses have been achieved 
in the field of GRBs. However, the energy mechanism of GRBs is still largely 
uncertain. Studies of beaming effects can potentially help to reveal this final
enigma. A good example is the possibility that the launch of a GRB jet might be 
related to the kick of a neutron star. This idea is proposed as early as in
1998 (Cen 1998), and has been discussed by a few authors (Dar \& Plaga 1999; 
Huang et al. 2003).

In a recent study, Huang et al. (2003) further suggested that the neutron
star should be a high speed one, with proper motion larger than $\sim 
1000$ km/s. In this frame-work, when a new-born neutron star is kicked, a high-speed
outflow should be launched into the opposite direction, whose energy can 
typically be $\sim 10^{52}$ ergs. The outflow may be composed of neutrinos and
anti-neutrinos initially. However, annihilation of neutrinos and anti-neutrinos 
can deposit a small portion ($\sim 10^{-3}$ --- $10^{-2}$) of its energy into 
an $e^{\pm}$ firecone. The isotropic equivalent energy of this firecone is 
$10^{50}$ --- $10^{54}$ ergs, depending on the energy deposition efficiency and 
the half opening angle. It thus can give birth to a beamed GRB successfully. 

This model, 
according to Huang et al.'s estimation, naturally meets many of the requirements
of GRB engines. For example, the deposited energy is enough for normal GRBs; 
the collimation is naturally guaranteed; the ultra-relativistic motion is 
reasonably produced; the observed connection between GRBs and supernovae is 
well explained; the duration of GRBs is consistent with the timescale of a typical
kick process; the event rate is satisfactory, i.e. consistent with the observed
GRB rate of $\sim$ 1 --- 3 per day; the model naturally produces the rapid 
variability in GRB light curves. Finally, it also explains the standard
energy reservoir hypothesis found by Frail et al. (2001).  

\section{Discussion and Conclusions}

In this article we introduce various beaming effects in GRBs. A convenient
way to calculate afterglows of beamed GRBs is introduced. Structures of GRB
jets are described, with the major attention being paid on the two-component 
model. We also discussed the possible existence of cylindrical jet in GRBs. 
The method of using orphan afterglow surveys to measure the beaming of GRB 
jets is discussed in some detail. It is shown that failed GRBs may play a
role in the process, and thus make the problem much more difficult. We also
addressed the possible connection between GRB jets and neutron star kicks.
We believe it is an interesting idea that the launch of a GRB jet may be 
associated with the kick of a high speed neutron star. 

Collimation is important in GRBs, since it provides important clues on the 
progenitors. Collimation can also be identified via effects other than 
those mentioned above. For example, optical afterglows from a jet can 
be significantly polarized, in principle up to tens of percents (Gruzinov
1999; Mitra 2000). In fact, polarization has already been observed in a 
few afterglows on the level of a few percents (Bersier et al. 2003). 
These observations strongly indicate that GRBs are collimated. However, 
such observations still cannot be directly used to measure the beaming 
angle. Radio afterglows in the very late phase can be used to estimate
the intrinsic kinetic energy of GRB remnant, and thus may provide 
information of beaming indirectly but independently. 

\acknowledgements
We thank the referee for useful comments and suggestions. 
This research was supported by the Special Funds for Major State
Basic Research Projects, the Foundation for the Author of National 
Excellent Doctoral Dissertation of P. R. China (Project No: 200125), 
Projects 10003001, 10233010 and 10221001 supported by NSFC, and an 
RGC grant of Hong Kong SAR.

\end{article}
\end{document}